\documentclass[a4paper,11pt]{article}
\usepackage{pos}
\usepackage{enumitem}
\usepackage{float}
\usepackage[numbers,sort&compress]{natbib}

\title{Emerging Narrow Resonance at 152 GeV}

\author*[a,b]{Srimoy Bhattacharya}
\author[a]{Mukesh Kumar}
\author[a]{Rachid Mazini}
\author[a,b]{Bruce Mellado}

\affiliation[a]{School of Physics and Institute for Collider Particle Physics, University of the Witwatersrand,\\
Johannesburg, Wits 2050, South Africa.}

\affiliation[b]{iThemba LABS, National Research Foundation,
PO Box 722, Somerset West 7129, South Africa.}

\emailAdd{srimoy.bhattacharya@cern.ch}
\emailAdd{mukesh.kumar@cern.ch}
\emailAdd{rachid.mazini@cern.ch}
\emailAdd{bmellado@mail.cern.ch}
\abstract{
The discovery of the Higgs boson at the LHC completed the Standard Model (SM), yet the possibility of additional scalars remains open, provided their contributions to electroweak symmetry breaking are sufficiently small. Recent analyses of LHC data have revealed statistically significant anomalies in multi-lepton final states - events characterized by multiple leptons, missing transverse energy, and ($b$-)jets. These anomalies provide intriguing hints of physics beyond the SM. In this work, we present the signature of growing excesses for a new scalar resonance with a mass of $152 \pm 1$ GeV, observed in the $\gamma\gamma$, $Z\gamma$, and $WW$ channels. The combined global significance reaches a level that points toward the growing signature of this resonance. The findings align with a simplified model in which a heavy scalar boson decays into two lighter scalars, thus providing a consistent framework explaining the observed multi-lepton anomalies.

These results significantly advance the search for new scalar bosons at the electroweak scale. Future investigations, including precision studies with upcoming HL-LHC data, will be crucial for confirming the nature of this resonance and exploring its implications for extending the SM.}

\FullConference{XXXII International Workshop on Deep Inelastic Scattering and Related Subjects (DIS2025)\\
24-28 March, 2025\\
Cape Town, South Africa\\}

\begin{document}
\maketitle

\section{Introduction}

The Standard Model (SM) of particle physics provides a comprehensive framework describing the fundamental particles and their interactions~\cite{ParticleDataGroup:2022pth}. The discovery of the Higgs boson in 2012 by ATLAS and CMS~\cite{Aad:2012tfa,Chatrchyan:2012ufa} confirmed the mechanism of electroweak symmetry breaking. Measurements of the $125$\,GeV boson show good agreement with SM expectations~\cite{Langford:2021osp}. Nonetheless, the SM leaves several key phenomena unexplained -- dark matter (DM), neutrino masses, and the baryon asymmetry -- while facing theoretical tensions such as the hierarchy problem. These issues suggest the existence of an extended scalar sector, which could also provide explanations for the observational evidence for DM and neutrino masses. Interestingly, over the past decade, several statistically significant anomalies have appeared in LHC data involving final states with multiple leptons (electrons and/or muons) in conjunction with missing energy and with or without ($b$-)jets ~\cite{Fischer:2021sqw,Crivellin:2023zui}, which indicate that the anomalies are not isolated but statistically consistent across channels and datasets. These ``multi-lepton anomalies’’ may hint at additional scalar bosons at the electroweak scale. 
Within the framework of a simplified model~\cite{vonBuddenbrock:2016rmr,Buddenbrock:2019tua}, the multi-lepton anomalies can be explained by the gluon-fusion production of a heavy scalar $H$ ($m_H \approx 270$\,GeV) decaying dominantly into two lighter scalars $S$ (one off-shell), each primarily decaying to $W$ bosons and $b$-jets. A mass of $150\pm5$\,GeV for $S$ was inferred from the invariant mass of electron–muon pairs in non-resonant $WW$ events~\cite{vonBuddenbrock:2017gvy}, motivating targeted searches for an associated production of narrow resonances around 150 GeV.

\section{Multi-lepton anomalies and model motivation}

\begin{figure}[t]
    \centering
    \includegraphics[width=0.6\linewidth]{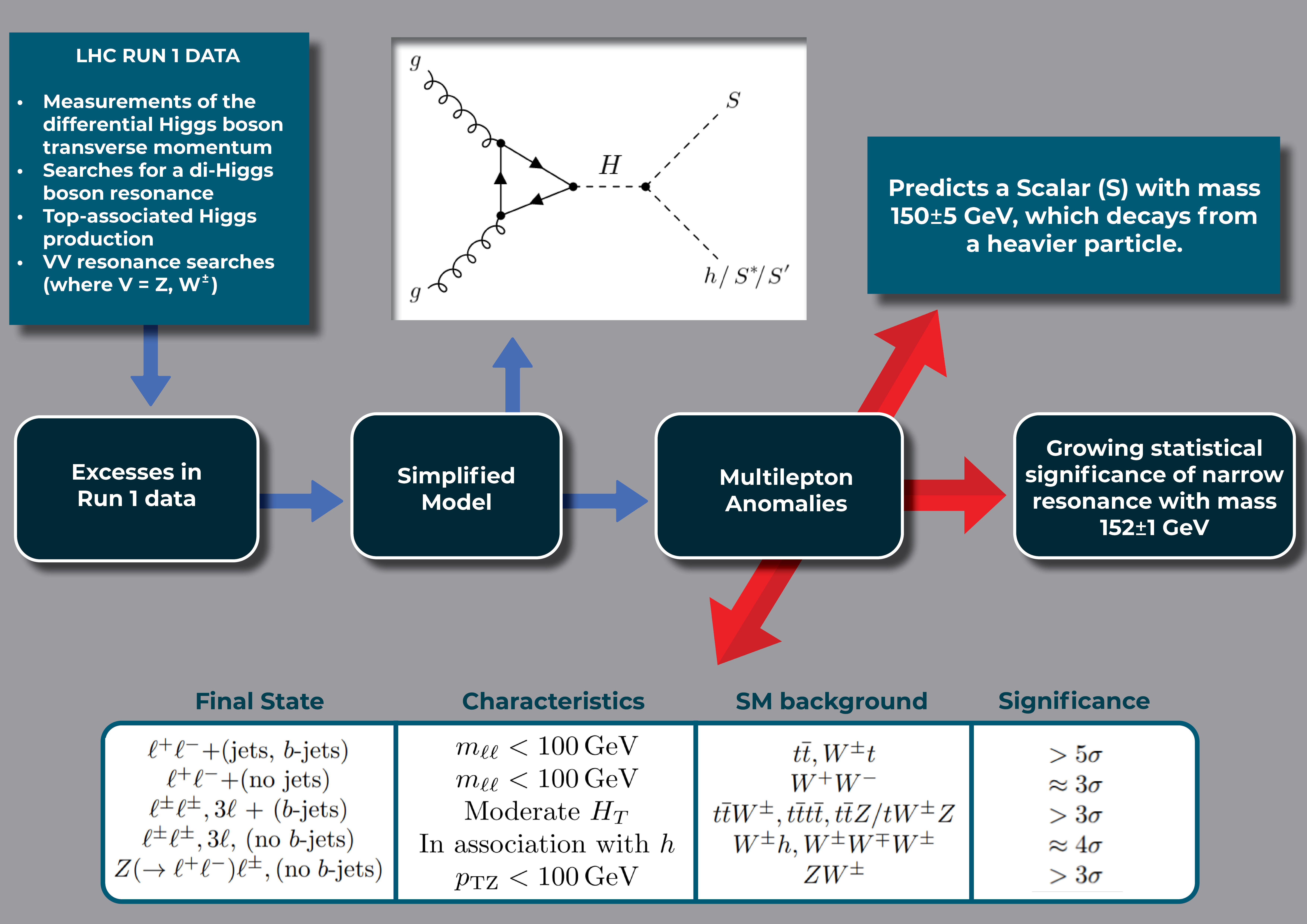}
    \caption{Timeline summarizing the evolution from early multi-lepton anomalies to the predicted narrow resonance near 150\,GeV~\cite{Crivellin:2023zui}.}
    \label{fig:timeline}
\end{figure}

Multi-lepton anomalies first appeared in LHC Run~1 data, including excesses in Higgs $p_T$ distributions~\cite{ATLAS:2014yga,ATLAS:2014xzb,CMS:2015qgt,CMS:2015hja}, di-Higgs searches~\cite{ATLAS:2015sxd,CMS:2014ipa,CMS:2015uzk,CMS:2014jkv}, and associated Higgs production with top quarks~\cite{ATLAS:2014ayi,ATLAS:2015xdt,ATLAS:2015utn,CMS:2014tll}. 
These features motivated a simplified model, as discussed earlier, predicting final states with $\ell^+\ell^-+E^T_\text{miss}$, $b$-jets, and photons. 
Subsequent studies~\cite{Buddenbrock:2019tua,Hernandez:2019geu} confirmed that these anomalies persisted in Run~2 data. 
More recently, excesses in triboson ($VVV$) final states with $V=W,Z$ have been reported, notably in $VVZ$, $tWZ$, and $WWW$ channels~\cite{ATLAS:2024nab,CMS:2025hlu,CMS:2020hjs,ATLAS:2022xnu,CMS:2025tre},which aligns with a Real Higgs Triplet model with hypercharge $Y = 0$~\cite{Ashanujjaman:2024lnr}.
These correlated features strengthen the case for an extended Higgs sector involving a new scalar around 150~GeV. 
Figure~\ref{fig:timeline} shows how anomalies accumulated chronologically, converging toward a predicted scalar around $150\pm5$~GeV.

A systematic search for a resonance near $m_S \approx 150 \pm 5~\text{GeV}$ was performed using data from ATLAS and CMS corresponding to proton--proton collisions at $\sqrt{s} = 13~\text{TeV}$. Rather than executing a broad, model-independent scan of all available final states, we adopted a targeted strategy guided by the simplified model introduced in Ref.~\cite{vonBuddenbrock:2016rmr}. This approach significantly reduces the ``look-elsewhere'' effect and potential biases associated with cherry-picking statistically interesting regions in a large dataset. The simplified model assumes that a heavy scalar $H$ with a mass of $ m_H\simeq 270~\text{GeV}$ is produced dominantly via gluon fusion and decays into two lighter scalars $S$, one of which can be off-shell ($ H\to SS^ {(*)}$). Each $S$ behaves as an SM-like scalar, decaying primarily to $W$ bosons or $b$-jets, while also allowing small branching ratios into photons and invisible states. This topology naturally predicts associated production of photons, leptons, and jets in the final state.

\section{Methodology}
\vspace{-0.2 cm}
The channels, predicted by the simplified model used to interpret the corresponding experimental results, are organized into three groups according to the time period of analysis and the evolution of the dataset:




\begin{itemize}[itemsep=0pt, topsep=1pt, parsep=0pt, partopsep=0pt,leftmargin= 10 pt]
    \item \textbf{Group 1 (2021):} Initial studies focused on diphoton ($\gamma\gamma$) and $Z\gamma$ final states with missing energy, $b$-jets, or additional leptons. The first combined analysis~\cite{Crivellin:2021ubm} yielded a global (local) significance of $3.9\sigma$ ($4.3\sigma$) near $151.5~\text{GeV}$.
    \item \textbf{Group 2 (2023):} Extended the search to include $\gamma\gamma+\geq(1\ell+1b\text{-jet})$, $\gamma\gamma+\geq4~\text{jets}$, and $WW^{*}+E_T^{\text{miss}}$ final states~\cite{ATLAS:2023omk,CMS:2022uhn,ATLAS:2022ooq}, further strengthening the case for a common underlying resonance.
    \item \textbf{Group 3 (2025):} The most recent results~\cite{ATLAS:2024lhu} explored the $\gamma\gamma+\tau$ and $\gamma\gamma+2(\ell,\tau)$ channels.
\end{itemize}

Each analysis utilized the sidebands of the SM Higgs searches to probe the invariant mass spectra between 140 and $155~\text{GeV}$. The local $p$-values were extracted from these spectra using the standard background models provided by the experimental collaborations, ensuring a consistent treatment of statistical uncertainties. For the $WW^{*}+E_T^{\text{miss}}$ final state, the results were derived by recasting the data from Ref.~\cite{Coloretti:2023wng} in the context of the simplified model. For each channel, we determined the best-fit signal yield and corresponding local $p$-value as a function of the hypothesized $S$-boson mass. The independent $p_i$ values were then combined using Fisher’s method~\cite{fisher1925}, which defines a global test statistic as:
\begin{equation}\textstyle
\chi^2_{2n} = -2 \sum_{i=1}^{n} \ln(p_i),
\end{equation}
where $n$ is the number of independent channels. The resulting statistic follows a $\chi^2$ distribution with $2n$ degrees of freedom. 
 In the final combination, five statistically independent channels were included:

 \begin{enumerate}[itemsep=0pt, topsep=2pt, parsep=0pt, partopsep=0pt]
     \item $S \to \gamma\gamma,\, Z\gamma$
     \item $S \to \gamma\gamma+\geq4~\text{jets}$
     \item $S \to \gamma\gamma+\geq(1\ell+1b\text{-jet})$
     \item $S \to WW^{*}+E_T^{\text{miss}}$
     \item Correlated $S \to \gamma\gamma+(1\tau,2\tau)$ channels.
 \end{enumerate}
The correlation between one- and two-$\tau$ final states was accounted for by including the ratio of signal efficiencies obtained from simulations of the $H \to S S^{(*)}$ process. A trial factor of approximately 3.5 was applied to correct for the look-elsewhere effect across the scanned mass range, consistent with the resolution of the diphoton analyses~\cite{Gross:2010qma}.
\vspace{-0.2 cm}
\section{Results and discussion}

\begin{figure}[t!]
    \centering
    \includegraphics[width=0.6\linewidth]{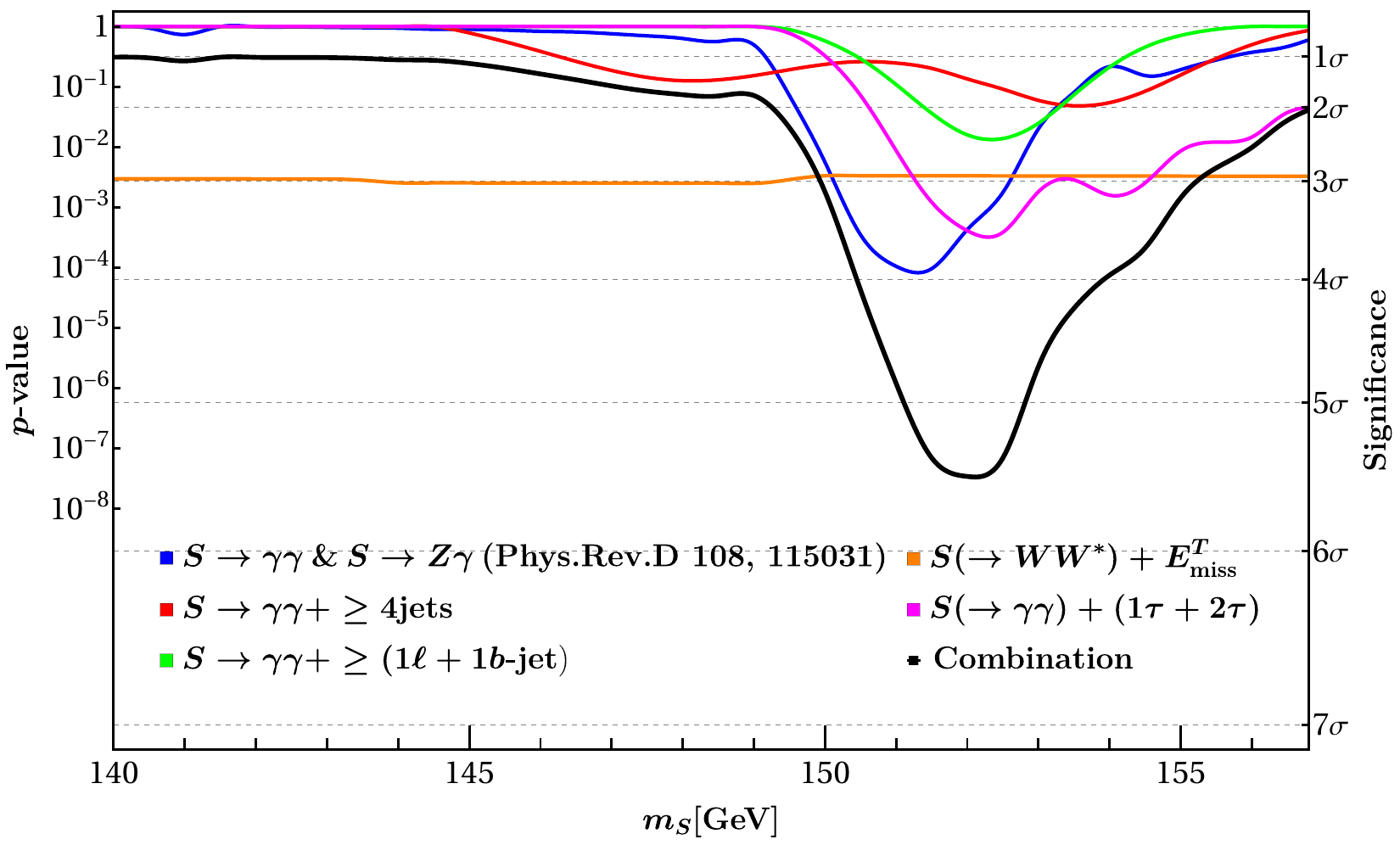}
    \caption{ Combined $p$-values as a function of the hypothesized scalar mass. The minimum occurs near $152~\text{GeV}$, corresponding to a global significance of $5.3\sigma$~\cite{Bhattacharya:2025rfr}.}
    \label{fig:pvalue}
\end{figure}

Using the methodology explained above, we first revisited the 2021 combination, incorporating additional $\gamma\gamma+b$-jet~\cite{ATLAS:2021ifb} and $Z(\to \ell^+\ell^-)\gamma$~\cite{CMS:2022ahq} channels. The inclusion of these datasets slightly reduced the global significance from $3.9\sigma$ to $3.5\sigma$ at $152~\text{GeV}$. When the new $WW^{*}$, $4j$, and $1\ell+b$-jet channels were included, the combined global (local) significance rose to $4.7\sigma$ ($4.9\sigma$)~\cite{Bhattacharya:2023lmu}. Finally, integrating the 2025 $\tau$-enriched channels led to a further increase, yielding a global (local) significance of $5.3\sigma$ ($5.5\sigma$) at $m_S = 152~\text{GeV}$, as shown in Figure.~\ref{fig:pvalue}~\cite{Bhattacharya:2025rfr}. The consistent excesses across all independent final states strongly indicate the presence of a narrow resonance, compatible with a SM-like scalar. 
The robustness of this excess was further tested against variations in background modeling and systematic uncertainties. These were found to have only a marginal impact on the significance, typically at the level of $ 0.1 \sigma$, comparable to the impact observed in the discovery of the 125 GeV Higgs boson~\cite{Aad:2012tfa}. This provides additional confidence that the observed structure is unlikely to originate from modeling artifacts.
The coherent appearance of signals across multiple independent final states and decay modes provides the strongest indication yet of a new scalar resonance at $152\pm1$~GeV. 
Confirming these states would mark a breakthrough beyond the SM, opening a new era in particle physics. 
A future high-precision $e^+e^-$ collider~\cite{FCC:2018evy,CEPCStudyGroup:2018ghi} will be essential to probe their properties and validate their theoretical interpretation.


\setlength{\bibsep}{0pt plus 0.3ex} 
\renewcommand{\bibfont}{\small}      

\bibliographystyle{unsrt}
\bibliography{sn-bibliography}


\end{document}